\documentclass[letter]{aa} 

\bibpunct{(}{)}{;}{a}{}{,} 

\usepackage[colorlinks = true,
            linkcolor = blue,
            urlcolor  = blue,
            citecolor = blue]{hyperref} 

\usepackage{graphicx}
\usepackage{txfonts}
\usepackage{tikz}
\usepackage{threeparttable}

\usepackage{lipsum}
\usepackage{subcaption}     
\usepackage{lscape}             
\usepackage{placeins}  
                                
\usepackage[
    starfontserif 
    ]{starfont}
\usepackage{amsmath}

\begin{document} 

   \title{The Neptunian ridge as a natural outcome\\of high-eccentricity tidal migration}

    
    \author{A.~Castro-Gonz\'{a}lez\inst{\ref{obs_geneva},\thanks{Corresponding author: \url{amadeo.castro-gonzalez@unige.ch}}}
    \and
    V. Bourrier\inst{\ref{obs_geneva}}
    \and
    D.~Ehrenreich\inst{\ref{obs_geneva}}
    \and
    D. J. Armstrong\inst{\ref{warwik}, \ref{warwik_2}}
    \and
    \\ 
    A.~C.~M.~Correia\inst{\ref{cfisuc-coimbra},\ref{obs_paris}}
    \and 
    M.~Lendl\inst{\ref{obs_geneva}}
    }
    
    \institute{Observatoire Astronomique de l’Université de Genève, Chemin Pegasi 51b, CH-1290 Versoix, Switzerland\label{obs_geneva}
    \and
    Centre for Exoplanets and Habitability, University of Warwick, Gibbet Hill Road, Coventry, CV4 7AL, UK\label{warwik}
    \and
    Department of Physics, University of Warwick, Gibbet Hill Road, Coventry, CV4 7AL, UK\label{warwik_2}
    \and
    CFisUC, Departamento de F\'isica, Universidade de Coimbra, 3004-516 Coimbra, Portugal\label{cfisuc-coimbra}
    \and
    LTE, Observatoire de Paris, Universit\'e PSL, Sorbonne Universit\'e, CNRS, 75014 Paris, France\label{obs_paris}
    }

\date{Received 23 February 2026 / Accepted 17 April 2026}

 
    \abstract
     {Recent occurrence-rate analyses have shown that the transition between the Neptunian desert and the savanna is not smooth but instead exhibits an overdensity of planets at orbital periods of $\simeq 3$-$6$~d, known as the Neptunian ridge.}
    {We confronted the high-eccentricity tidal migration (HEM) scenario with the Neptunian desert--ridge--savanna landscape.}
    {We mapped the HEM tidal survival constraints onto the period--radius plane using empirically inferred mass--radius relations, and provide an additional consistency check by projecting the tidal survival boundary onto the period--density plane.}
    {The HEM tidal survival formalism reproduces the slope of the Neptunian desert boundary across the sub-Neptune to super-Neptune/sub-Saturn regime (1.8\,$\rm R_\oplus$ $\lesssim R_{\rm p} \lesssim$ 6\,$\rm R_\oplus$), with a single representative value of the tidal encounter parameter setting the overall period offset. In the Jovian regime, the empirical boundary geometry remains broadly consistent with the tidal survival limit, although additional effects such as radius inflation and orbital decay may account for residual deviations. Incorporating the empirically inferred density dispersion transforms the disruption limit into a finite tidal survival band that traces the Neptunian ridge. Because tidal dissipation rises steeply towards the disruption threshold, HEM survivors are expected to circularise just beyond this limit, thereby clustering within the survival band and naturally generating the ridge overdensity. In the period--density plane, the observed population follows the predicted density-dependent tidal survival and clustering pattern, and the high-density ridge planets exhibit a persistent concentration near $\rho_{\rm p}\simeq1.7\,\mathrm{g\,cm^{-3}}$, whose origin warrants further investigation.}
    {High-eccentricity tidal migration provides a self-consistent physical explanation for the origin of the Neptunian ridge and the geometry of the desert boundary.}
   \maketitle
%
 \section{Introduction}
\label{sec:intro}

Soon after the first discoveries, a deficit of Neptunian planets at very short orbital periods was identified \citep[e.g.][]{Mazeh2005,Mazeh2016,Lecavelier2007}, a feature commonly referred to as the Neptunian desert. This deficit reflects a genuine scarcity of planets rather than an observational bias. Occurrence-rate analyses have shown that the desert remains highly underpopulated even when survey incompleteness is taken into account \citep{CastroGonzalez2024a,2026MNRAS.546ag022C}. However, its origin remains an open problem \citep[][]{Matsakos2016,Bourrier2018Natur,OwenLai2018}.

Atmospheric escape can erode gas-rich planets at short orbital separations and has been invoked as a plausible contributor to the desert \citep[][]{OwenLai2018}. In parallel, high-eccentricity tidal migration \citep[HEM; e.g.][]{FordRasio2008,2011CeMDA.111..105C} has been widely discussed as a mechanism shaping the desert. If outer perturbers drive planets to sufficiently small periastrons, tidal dissipation can lead either to disruption or to rapid circularisation onto compact final orbits, with characteristic periods comparable to those of the desert edge \citep[e.g.][]{2013ApJ...762...37L,Mazeh2016,Matsakos2016}. Today, it is increasingly thought that an interplay between these mechanisms shapes the properties of close-in Neptunes \citep[e.g.][]{Bourrier2018Natur,OwenLai2018}, with contributions from tidal inflation and orbital decay \citep[e.g.][]{HallattMillholland2026,Hallatt2026}.

Recently, the observational picture of the `exo-Neptunian landscape' has sharpened. Using occurrence-rate calculations to correct for survey biases, \citet{CastroGonzalez2024a} redefined the boundaries of the Neptunian desert and show that it is bordered by an overdensity of planets concentrated at $P_{\rm orb}\simeq 3.2$-$5.7$~d, which separates the desert from the more sparsely populated, longer-period Neptunian savanna \citep{Bourrier2023}. This feature, referred to as the Neptunian ridge, resembles the hot-Jupiter pile-up, suggesting that the closest-in Neptune- to Jupiter-size planets share related evolutionary pathways \citep[e.g.][]{CastroGonzalez2024a}. The origin of the pile-up remains debated, but HEM is widely considered to play a key role (see \citealt{Dawson2018} and references therein). Recent observational evidence points to a similarly important contribution in the ridge \citep[e.g.][]{Correia2020,Bourrier2023,Bourrier2025,Doyle2025,Vissapragada2025,2025AJ....169..212H,Yee2025,2026ApJ...998..324H}.

In this work we revisit the HEM scenario in light of the updated, bias-corrected geometry of the Neptunian desert, ridge, and savanna inferred by \citet{CastroGonzalez2024a}. 

\section{Tidal survival and circularisation}
\label{sec:tidal}

In HEM scenarios, planets are driven onto highly eccentric orbits through mechanisms such as secular perturbations and planet--planet scattering. Depending on their periastron distance, they can be disrupted, collide with the star, be ejected, or survive and circularise. Their fate is governed by tidal interactions near periastron, which impose a fundamental constraint on the resulting population of close-in survivors.

Planets that approach their host stars too closely are tidally stripped and can be fully disrupted. This defines a minimum periastron distance set by the tidal radius \citep{Roche1849},
\begin{equation}
r_{\mathrm{tide}} = \eta R_{\mathrm{p}}\left(\frac{M_\star}{M_{\mathrm{p}}}\right)^{1/3},
\label{eq:rtide}
\end{equation}
where $\eta$ is a dimensionless encounter parameter that depends on the planetary internal structure and on the hydrodynamics of the tidal interaction. This condition translates into a minimum circularised orbital period after tidal dissipation. Because the limit depends on the planet's bulk density, it can be mapped onto the period--radius plane through a mass--radius relation, defining a tidal survival boundary, $P_{\rm tide}(R_{\rm p})$, for planets delivered through HEM \citep{Matsakos2016}.

Survival alone does not guarantee compact post-HEM orbits. Tidal circularisation is highly sensitive to periastron distance, so planets that avoid very close passages may survive but fail to circularise on system timescales \citep[e.g.][]{FabryckyTremaine2007}. Tidal evolution therefore operates most efficiently in a narrow periastron range near the disruption threshold.

Appendix~\ref{app:A1} gives the explicit expressions used to evaluate $P_{\rm tide}(R_{\rm p})$, and Appendices~\ref{app:A2} and \ref{app:A3} show illustrative circularisation scalings for the Lidov-Kozai and secular-chaos channels. Other pathways, such as planet--planet scattering, are expected to show similar behaviour, since the outcome is set mainly by the tidal physics rather than by the specific excitation mechanism.

\section{A tidal origin for the Neptunian ridge}
\label{sec:connect}

\begin{figure*}
    \centering
    \includegraphics[width=0.94\textwidth]{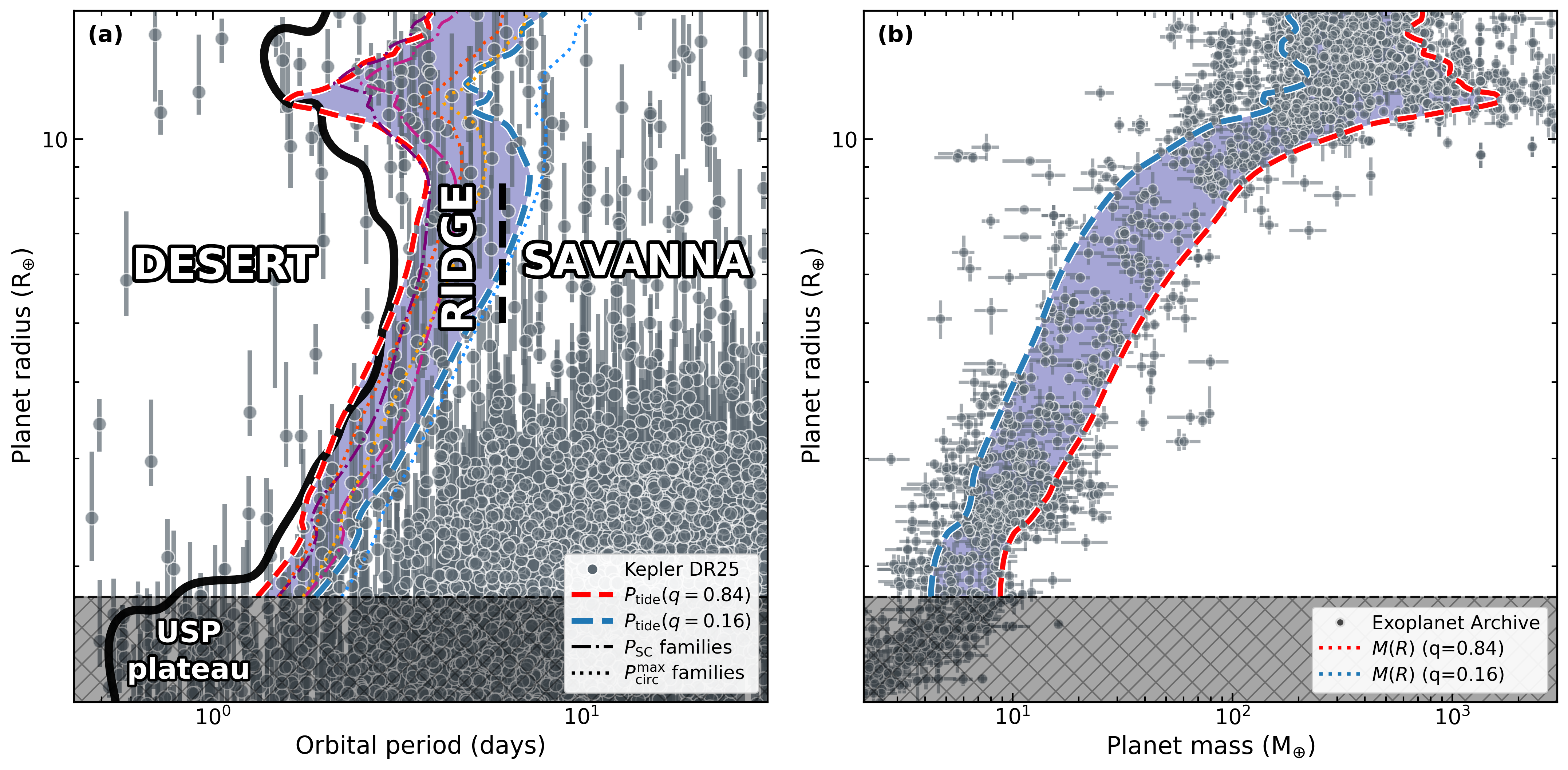}
    \caption{Confrontation between tidal survival limits and the close-in planet population. \textit{(a)} Period--radius plane. The solid black curve traces the Neptunian desert boundary, and the vertical dashed line marks the ridge limit \citep{CastroGonzalez2024a}. Dashed red and blue curves show tidal survival boundaries $P_{\rm tide}(R_{\rm p};q)$ for the $q=0.84$ and $q=0.16$ mass--radius envelopes, adopting $\eta=4.5$, and the shaded region indicates the survival band. Dotted curves show representative $P_{\rm circ}^{\max}$ tracks with $Q'_{\rm p}=10^5$ and $M_\star=1$\,$\rm M_\odot$ \citep{OwenLai2018}: orange and red correspond to $t_{\rm sys}=8$ and $4$~Gyr with $a_0=1.5$~au along the $q=0.84$ envelope, and blue to $t_{\rm sys}=8$~Gyr with $a_0=1.0$~au along the $q=0.16$ envelope. Dash-dotted dark magenta and pink curves show illustrative secular-chaos circularisation curves \citep[e.g.][]{Matsakos2016}, computed using $M(R_{\rm p};q=0.5)$ and perturbing-companion masses of $0.7$ and $0.4$\,$\rm M_{\rm J}$, respectively. \textit{(b)} Mass--radius plane with the quantile envelopes $M(R_{\rm p};q)$.}
    \label{fig:tidal_boundaries}
\end{figure*}

We confronted the HEM tidal constraints with the distribution of short-period planets, focusing on the empirically determined, bias-corrected boundaries of the Neptunian desert and ridge (Fig.~\ref{fig:tidal_boundaries}; \citealt{CastroGonzalez2024a}). A key feature of the updated landscape is the flattening of the lower desert boundary at $R_{\rm p}\simeq 1.8$\,$\rm R_\oplus$ over $P_{\rm orb}\simeq0.5$-$1.2$~d, which we refer to as the ultra-short-period plateau. We restricted our analysis to $R_{\rm p}>1.8$\,$\rm R_\oplus$ since rocky planets possess significant material strength. Moreover, the shortest-period planets systematically approach the disruption limit \citep{CastroGonzalez2024a,CastroGonzalez2025}, likely requiring orbital decay or alternative migration channels.

\subsection{Mass--radius envelopes and tidal survival limits}

To map the theoretical tidal limits onto the $(P,R_{\rm p})$ plane, a mass--radius relation is required. We constructed this relation empirically using planets with measured masses from the NASA Exoplanet Archive, retaining only objects with sufficiently precise measurements ($\sigma_{M_{\rm p}}/M_{\rm p} \leq 20\%$ and $\sigma_{R_{\rm p}}/R_{\rm p} \leq 10\%$) to ensure fractional density uncertainties $\lesssim 35\%$ (for reference, the $1\sigma$ density scatter is $\simeq2.3\,\mathrm{g\,cm^{-3}}$ in the $10$-$30$\,$\rm M_\oplus$ regime).

At fixed radius, envelopes of $\log M$ are estimated using quantile statistics, yielding relations $M_{\rm p}(R_{\rm p};q)$, where $q$ denotes the cumulative quantile level. We adopted the quantile pair $q=0.8413$ and $q=0.1587$, which correspond to the $\pm1\sigma$ levels of a normal distribution. These envelopes represent the high- and low-density ends of the gas-rich planet population. Together, they define a finite tidal survival band in the period--radius plane,
\begin{equation}
P_{\rm tide}(q=0.8413) < P \leq P_{\rm tide}(q=0.1587),
\label{eq:band}
\end{equation}
which captures the range of the shortest post-HEM orbits reachable for the bulk of systems without tidal disruption once the dispersion in density is taken into account.\footnote{Adopting more extreme quantiles (e.g. $q = 0.05$ and $q = 0.95$) yields broader and less smooth envelopes but does not alter the conclusions.}

\subsection{A coupled HEM origin of the ridge and desert boundary}
\label{sec:HEM_origin}

The slope of the desert boundary from the sub-Neptune to the super-Neptune/sub-Saturn regime (1.8\,$\rm R_\oplus$ $\lesssim R_{\rm p} \lesssim$ 6\,$\rm R_\oplus$) follows directly from the HEM tidal survival limit when evaluated along the high-density envelope, $P_{\rm tide}(q=0.84)$. This agreement is not enforced by construction but instead emerges from the tidal survival criterion itself. Adopting a single representative value of the tidal encounter parameter, $\eta = 4.5$, which sets the overall period offset, the predicted boundary reproduces the bias-corrected desert edge over this full radius range. We note that $\eta$ should not be interpreted here as a universal disruption constant but rather as an effective parameter for the onset of strong tidal disruption within the considered framework. In particular, the derived value is tied to the definition of the desert, the mass--radius relation, and the quantile envelopes. Still, the effective $\eta$ inferred here is of the same order of magnitude as those reported in previous works, including the $\eta = 2.7$ from hydrodynamic calculations for Jupiter-like planets \citep{Guillochon2011} and the $\eta = 3.5$ from a related HEM study \citep{Matsakos2016}.

In the Jovian regime, additional effects such as radius inflation or post-circularisation orbital decay are likely required to account for residual deviations from the measured boundary, although the overall boundary shape and slope transitions remain broadly consistent with the tidal survival limit. Because a single value of $\eta$ closely reproduces the overall desert geometry, we adopted it consistently to construct the tidal survival band. Remarkably, this band traces the period interval of the Neptunian ridge, $3 \lesssim P_{\rm orb} \lesssim 6$~d (Fig.~\ref{fig:tidal_boundaries}). The ridge therefore emerges as the natural locus of the shortest-period HEM survivors once the observed density dispersion is taken into account. We note that this interpretation assumes that the density distribution relevant for HEM was broadly similar to that observed today. If some planets migrated at younger ages or were temporarily inflated during HEM, the low-density tail and thus the tidal survival band could have been broader, which would also affect the value of $\eta$.

Tidal dissipation increases very steeply towards small periastron distances, so efficient circularisation is concentrated among trajectories that pass close to the disruption threshold. As a consequence, successful HEM outcomes typically accumulate just beyond the survival limit \citep[e.g.][]{FabryckyTremaine2007,Giacalone2017,OwenLai2018}. This concentration is intrinsically stronger for lower-density planets, whose larger disruption radii shift the survival boundary to longer periods. Therefore, the combination of the steep tidal dependence on periastron distance and the finite density dispersion naturally confines surviving HEM outcomes to a narrow period interval that traces the Neptunian ridge. Figure~\ref{fig:tidal_boundaries} also shows representative maximum circularisation periods for the Lidov-Kozai channel and illustrative secular-chaos circularisation curves. While these first-order estimates are model-dependent and intended only as qualitative guides, their agreement with the desert--ridge structure supports the notion that HEM plays a major role in setting its geometry.

At larger radii, the same tidal framework predicts an overdensity of HEM-migrated planets in the hot-Jupiter pile-up (Fig.~\ref{fig:tidal_boundaries}). At smaller radii, the HEM-accessible region is embedded within the high-occurrence sub-Neptune population, making any overdensity harder to isolate and requiring additional observables to determine a HEM contribution.

\subsection{Tidal survival and clustering in the period--density plane}

\begin{figure}
    \centering
    \includegraphics[width=0.46\textwidth]{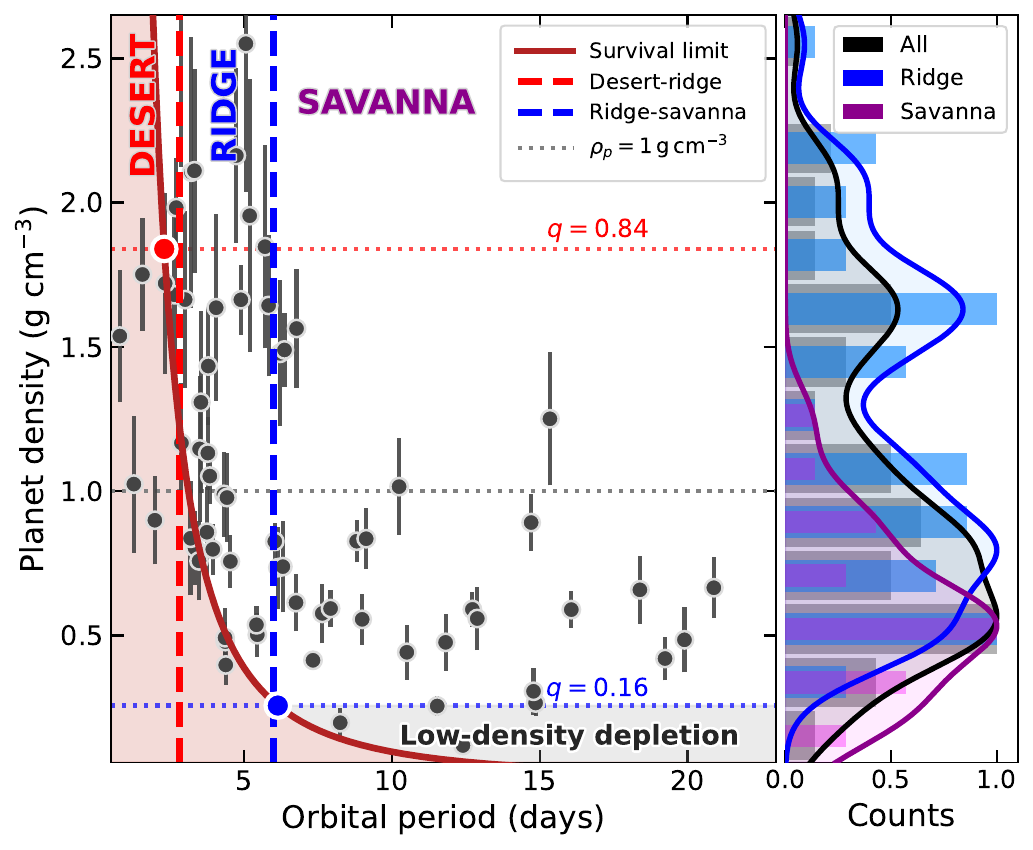}
   \caption{Left: Period--density distribution of close-in Neptunian planets with precise masses and radii ($4\leq R_{\rm p}/R_\oplus\leq8.5$). The red curve shows the tidal survival limit adopting $\eta=4.0$. Vertical dashed lines mark the desert--ridge and ridge--savanna boundaries \citep{CastroGonzalez2024a}, and horizontal dotted blue and red lines indicate the low- and high-density quantiles and their intersections with the tidal limit. The grey-shaded region highlights the low-density depletion, which effectively limits typical HEM outcomes to the ridge interval. Right: Normalised density distribution, defining the savanna as $P_{\rm orb}>8$~d.}
    \label{fig:density_neptunes}
\end{figure}

We further tested the HEM tidal survival formalism by projecting the close-in Neptunian population onto the period--density plane. Because the tidal survival condition depends on bulk density, it translates into a simple physical boundary in this space. For this analysis, we adopted $\eta = 4.0$, chosen so that the bulk of the planets lie outside the disruption region (Fig.~\ref{fig:density_neptunes}). This value is slightly lower than the $\eta = 4.5$ adopted in Sect.~\ref{sec:HEM_origin}, which is not unexpected since the two estimates rely on different assumptions regarding the desert boundary and the mass--radius distribution. 

The observed distribution shows two significant features previously reported in the literature: a group of ridge Neptunes are systematically denser than savanna planets, with a characteristic transition near $1\,\mathrm{g\,cm^{-3}}$ \citep{CastroGonzalez2024b}, and the lower envelope of the density distribution rises towards shorter periods (the density `brink’; \citealt{Bourrier2025}). In Fig.~\ref{fig:density_neptunes}, this lower envelope closely traces the predicted tidal survival limit. This agreement supports a scenario in which tidal disruption sets the minimum density required for planets to survive at a given orbital period under HEM. At fixed radius, higher-density planets can survive smaller periastron distances and reach shorter post-HEM periods, whereas lower-density planets are preferentially removed and are therefore absent at the shortest periods. 

The same diagram also naturally explains the accumulation of HEM survivors in the ridge. The tidal survival curve intersects the low-density quantile of the Neptunian population at $\simeq$\,6~d, closely matching the ridge--savanna boundary (Fig.~\ref{fig:density_neptunes}). This result ultimately reflects the fact that the density distribution of Neptunian planets does not extend to sufficiently low values to effectively shift the minimum tidal survival period to orbital periods longer than the ridge interval. This depletion of very low-density planets is highlighted in Fig.~\ref{fig:density_neptunes}. Overall, this density-limited survival window naturally explains both the confinement of HEM survivors to the ridge and the observed density gradient towards shorter periods.

A key open question is the origin of the dense Neptunes in the ridge. \citet{CastroGonzalez2024b} suggest that the broad density spread reflects either intrinsically different internal structures, possibly linked to multiple formation pathways, or evolutionary effects such as atmospheric erosion. Within the HEM framework, another possibility is partial tidal stripping of initially low-density ridge planets during close periastron passages. Interestingly, in the right-hand panel of Fig.~\ref{fig:density_neptunes} we identify a non-Gaussian density distribution, with a main peak at $\rho_{\rm p}\simeq0.6$-$0.8\,\mathrm{g\,cm^{-3}}$ and a secondary maximum near $\rho_{\rm p}\simeq1.7\,\mathrm{g\,cm^{-3}}$. This feature is most evident in the ridge subsample and appears robust to the analyses presented in Appendix~\ref{app:clustering}. We therefore interpret it as a real density concentration associated with the ridge, whose origin deserves further investigation.

\section{Discussion}
\label{sec:disc}

The results presented in this work support a unified interpretation of the Neptunian ridge and the adjacent desert boundary within the HEM scenario, sharpening earlier HEM-based explanations. \citet{Matsakos2016} showed that tidal-survival and secular-chaos curves can reproduce the basic desert geometry when combined with an empirical mass--radius relation. In a subsequent study, \citet{OwenLai2018} showed that post-HEM circularisation typically occurs only slightly beyond the disruption boundary, implying a narrow range of final periods. Our key differences are that we confronted the HEM framework with the updated, bias-corrected desert--ridge boundaries and explicitly incorporated the observed dispersion in the mass--radius relation, which broadens the survival limit into a finite tidal survival band whose full extent naturally spans the Neptunian ridge and the hot-Jupiter pile-up. We note that recent models suggest that the HEM-driven circularisation of ridge Neptunes often causes runaway tidal inflation unless the planets are sufficiently metal-rich or tidal dissipation occurs mainly in their outer layers \citep{Hallatt2026}. Atmospheric survival may therefore limit which HEM products remain in the ridge.

Independent observational trends further support a HEM origin for the Neptunian ridge overdensity. The excited eccentricities, polar orbits, and evidence of atmospheric erosion in several ridge planets are more consistent with a late, dynamically excited origin than with early, smooth disk-driven migration \citep[e.g.][]{Correia2020,Bourrier2023,Bourrier2025}. Host-star properties point in the same direction: ridge and desert Neptunes preferentially orbit metal-rich stars relative to savanna planets \citep{Vissapragada2025}, consistent with a higher occurrence of massive companions capable of triggering HEM. 

Finally, although HEM reproduces the desert--ridge geometry, it does not readily explain the savanna, which is mostly populated by nearly circular planets \citep{Correia2020}. In this context, \citet{Bourrier2025} propose that low-density Neptunes migrating early through the disk are strongly eroded at short periods, while those left farther out populate the savanna. By contrast, planets delivered later to the ridge through HEM would be less affected by early high-energy irradiation and could therefore survive at short periods \citep[][]{Bourrier2023,Attia2021}.

\section{Conclusions}

We revisited the HEM scenario using the bias-corrected geometry of the Neptunian desert--ridge structure. The tidal disruption condition naturally reproduces the desert boundary, and the density dispersion broadens the disruption limit into a finite tidal survival band that traces the ridge. Since tidal dissipation increases steeply towards the disruption threshold, HEM survivors are expected to circularise just beyond this limit and thus cluster within the survival band, making the coupled desert boundary and ridge overdensity a natural outcome of HEM. The period--density plane provides an independent consistency check and reveals a non-Gaussian density distribution with a robust local maximum near $\rho_{\rm p}\simeq1.7\,\mathrm{g\,cm^{-3}}$ that is primarily associated with ridge planets and whose origin warrants further investigation.

\begin{acknowledgements}
We thank the anonymous referee for a very careful, constructive, and insightful report, which significantly improved the clarity and overall quality of this work.
This work has been carried out within the framework of the NCCR PlanetS supported by the Swiss National Science Foundation under grant 51NF40\_205606.
ACMC acknowledges support from the FCT, Portugal, through the CFisUC project UID/04564/2025, with DOI identifier 10.54499/UID/04564/2025.
This work has made use of the NASA Exoplanet Archive \citep{Christiansen2025} and the \textit{Kepler} DR25 catalogue \citep{Thompson2018}.
This research made use of \texttt{nep-des} (available in \url{https://github.com/castro-gzlz/nep-des}).

\end{acknowledgements}

%
%

\bibliographystyle{aa} 
\bibliography{references} 
\begin{appendix}

\section{Tidal survival and circularisation}
\subsection{Tidal disruption and minimum circularisation period}
\label{app:A1}

For $M_{\mathrm{p}} \ll M_\star$, a planet of mass $M_{\mathrm{p}}$ and radius $R_{\mathrm{p}}$ orbiting a star of mass $M_\star$ undergoes significant tidal stripping or disruption if its periastron approaches the characteristic tidal radius \citep{Roche1849,Chandrasekhar1963},
\begin{equation}
r_{\mathrm{tide}} = \eta \, R_{\mathrm{p}} \left(\frac{M_\star}{M_{\mathrm{p}}}\right)^{1/3},
\label{eq:rtide_app}
\end{equation}
where the dimensionless parameter $\eta$ encapsulates the effects of planetary internal structure and the hydrodynamics of the encounter \citep[e.g.][]{Guillochon2011}. Following previous related studies, as a first approximation we assume a single value of $\eta$ over the density range of gas-rich planets from sub-Neptunes to Jupiters \citep[e.g.][]{Matsakos2016,OwenLai2018}.

In HEM scenarios, planets are driven to highly eccentric orbits ($e \rightarrow 1$). 
If angular momentum is conserved, the final circularised semi-major axis is \citep{Ford2006}
\begin{equation}
a_{\mathrm{F}} \simeq 2\, r_{\mathrm{p}},
\end{equation}
where $r_{\mathrm{p}}$ is the periastron distance. Requiring survival against tidal disruption implies
\begin{equation}
a_{\mathrm{F}} \geq 2\, r_{\mathrm{tide}}.
\label{eq:af_rtide_app}
\end{equation}
Using Kepler’s third law, this condition translates into a minimum circularisation period,
\begin{equation}
P_{\mathrm{F}} \geq
2\pi \left(\frac{(2\,r_{\mathrm{tide}})^3}{G M_\star}\right)^{1/2}.
\label{eq:pf_app}
\end{equation}
Substituting Eq.~(\ref{eq:rtide_app}) into Eq.~(\ref{eq:pf_app}) yields
\begin{equation}
P_{\mathrm{F}} \geq
2\pi \left(\frac{(2\,\eta\,R_{\mathrm{p}})^3}{G\,M_{\mathrm{p}}}\right)^{1/2}
\propto
\left(\frac{R_{\mathrm{p}}^3}{M_{\mathrm{p}}}\right)^{1/2}
\propto \rho_{\mathrm{p}}^{-1/2},
\label{eq:A5}
\end{equation}
showing that the tidal survival boundary is controlled primarily by the planet’s mean density and is independent of stellar mass. Once a mass--radius relation is specified, this condition defines a tidal survival boundary $P_{\rm tide}(R_{\rm p})$ in the period--radius plane.

\subsection{Lidov-Kozai circularisation}
\label{app:A2}

Following \citet{OwenLai2018}, we estimate the maximum orbital separation from which tidal dissipation can circularise a highly eccentric orbit within a system age $t_{\rm sys}$. This estimate is based on the equilibrium-tide formalism in the weak-friction approximation, appropriate for HEM driven by secular perturbations such as Lidov-Kozai cycles \citep[e.g.][]{FabryckyTremaine2007,Anderson2016}.

Requiring that tidal dissipation circularises the orbit within $t_{\rm sys}$ yields an upper limit on the circularised semi-major axis,
\begin{equation}
a_{\rm F} \leq a_{\rm circ},
\end{equation}
with
\begin{equation}
\begin{aligned}
a_{\rm circ} &=
0.05~\mathrm{au}\;
\left(\frac{t_{\rm sys}}{1~\mathrm{Gyr}}\right)^{1/7}
\left(\frac{Q'_p}{10^5}\right)^{-1/7}
\left(\frac{M_\star}{M_\odot}\right)^{2/7} \\
&\quad \times
\left(\frac{a_0}{1~\mathrm{au}}\right)^{-1/7}
\left(\frac{M_{\rm p}}{M_{\rm J}}\right)^{-1/7}
\left(\frac{R_{\rm p}}{R_{\rm J}}\right)^{5/7},
\end{aligned}
\label{eq:acirc}
\end{equation}
where $Q'_p$ is the effective planetary tidal quality factor and $a_0$ is the characteristic initial semi-major axis.

The corresponding maximum orbital period reachable by tidal circularisation is obtained via Kepler’s third law,
\begin{equation}
P_{\rm circ}^{\max} =
2\pi \left(\frac{a_{\rm circ}^3}{G M_\star}\right)^{1/2}.
\label{eq:pcirc}
\end{equation}

\noindent Equation~(\ref{eq:pcirc}) provides a first-order estimate of the maximum final periods reachable on a timescale $t_{\rm sys}$.

\subsection{Secular-chaos circularisation}
\label{app:A3}

Following \citet{Matsakos2016}, we consider the secular-chaos circularisation scaling from \citet{WuLithwick2011}. The characteristic post-circularisation semi-major axis is written as
\begin{equation}
\begin{aligned}
a_{\rm c,SC} &=
0.03~\mathrm{au}\;
\left(\frac{\alpha}{1/6}\right)^{-3/5}
\left(\frac{M_\star}{M_\odot}\right)^{2/5} \\
&\quad \times
\left(\frac{M_{\rm pert}}{M_{\rm J}}\right)^{-1/5}
\left(\frac{M_{\rm p}}{M_{\rm J}}\right)^{-2/5}
\left(\frac{R_{\rm p}}{R_{\rm J}}\right),
\end{aligned}
\label{eq:asc}
\end{equation}
where $\alpha$ is a dimensionless normalisation parameter and $M_{\rm pert}$ is the mass of the perturbing companion. As in \citet{Matsakos2016}, we adopt $\alpha=1/6$, which sets the reference normalisation of the scaling.
The corresponding orbital period is then obtained from Kepler's third law,
\begin{equation}
P_{\rm SC} =
2\pi \left(\frac{a_{\rm c,SC}^3}{G M_\star}\right)^{1/2}.
\label{eq:psc}
\end{equation}

\section{Density clustering diagnostics}
\label{app:clustering}

We quantify the evidence for a persistent density feature near $\rho_{\rm p}\simeq1.7\,\mathrm{g\,cm^{-3}}$ using complementary diagnostics applied consistently to the same parent sample. We select planets with measured masses and radii from the NASA Exoplanet Archive and apply the adopted radius range ($4.0 \le R_{\rm p}/R_\oplus \le 8.5$) and precision cuts ($\sigma_{M_{\rm p}}/M_{\rm p} \le 20\%$, $\sigma_{R_{\rm p}}/R_{\rm p} \le 10\%$). Densities are computed as $\rho_{\rm p}=(M_{\rm p}/R_{\rm p}^3)\,\rho_\oplus$, with uncertainties propagated as
\[
\frac{\sigma_{\rho}}{\rho_{\rm p}}\simeq\sqrt{\left(\frac{\sigma_{M_{\rm p}}}{M_{\rm p}}\right)^2 + \left(3\frac{\sigma_{R_{\rm p}}}{R_{\rm p}}\right)^2}.
\]
Measurement uncertainties are propagated either analytically (for window occupancies) or via Monte Carlo resampling of $\rho_{\rm p}$ from per-planet Gaussian error models when assessing the stability of the diagnostics.

\subsection{KDE peak stability near $\rho_{\rm p}\simeq1.7\,\mathrm{g\,cm^{-3}}$}

We analyse the density distribution using kernel-density estimation (KDE) in $\rho_{\rm p}$ with a Gaussian kernel. To avoid conclusions driven by a particular smoothing choice, we explore a grid of bandwidths spanning $h=0.06$-$0.22\,\mathrm{g\,cm^{-3}}$ (17 values), from under- to over-smoothed regimes. For each bandwidth, we evaluate whether the KDE profile exhibits a local maximum within a fixed narrow tolerance of the target density $\rho_{\rm p}=1.7\,\mathrm{g\,cm^{-3}}$.

In the total sample, a local KDE maximum near $1.7\,\mathrm{g\,cm^{-3}}$ is recovered for 94\% of the tested bandwidth choices. In the ridge subsample, the same peak is recovered for 100\% of the bandwidth grid. When propagating measurement uncertainties through Monte Carlo resampling, the fraction of realisations retaining a local maximum at $1.7\,\mathrm{g\,cm^{-3}}$ is 0.58 for the total sample and 0.70 for the ridge sample. This bandwidth- and uncertainty-level robustness indicates that the $\simeq1.7\,\mathrm{g\,cm^{-3}}$ feature is unlikely to be an artefact of smoothing or measurement noise.

\subsection{Gaussian-mixture evidence against a unimodal null}

We fit one- and two-component Gaussian mixture models to the $\rho_{\rm p}$ distribution and compare them using the Bayesian information criterion (BIC), computing $\Delta{\rm BIC}={\rm BIC}_{K=1}-{\rm BIC}_{K=2}$. For the total sample, we obtain $\Delta{\rm BIC}=15.7$, while for the ridge subsample $\Delta{\rm BIC}=3.35$.

To assess whether these values can arise under a unimodal distribution, we perform a parametric bootstrap under a single-Gaussian null and recompute $\Delta{\rm BIC}$ for each realisation. Under this unimodal null, the probability of obtaining $\Delta{\rm BIC}$ at least as large as observed is $p=4.0\times10^{-4}$ for the total sample and $p=3.6\times10^{-3}$ for the ridge sample. Consistently, the unimodal null yields ${\rm P}(\Delta{\rm BIC}>0)\simeq0.028$ (total) and $0.009$ (ridge).

When uncertainties are propagated via Monte Carlo resampling, the median $\Delta{\rm BIC}$ remains positive for the total sample (median $\simeq12.5$, 68\% interval [7.2,\,18.6]), while for the ridge subsample the distribution of $\Delta{\rm BIC}$ spans zero (median $\simeq-0.1$, 68\% interval [-4.1,\,4.4]), indicating weaker formal evidence for a fully separated bimodal decomposition in this subsample.

\subsection{Local bump strength and window occupancy}

We further quantify the prominence of the $\simeq1.7\,\mathrm{g\,cm^{-3}}$ feature using complementary local diagnostics. We define a `1.7-peak window' of $\rho_{\rm p}\in[1.45,\,2.05]\,\mathrm{g\,cm^{-3}}$. A parametric local-excess test under the fitted unimodal null yields $p=0.236$ (total) and $p=0.112$ (ridge).

Accounting analytically for measurement uncertainties by summing the per-planet Gaussian probabilities of falling inside the window yields an uncertainty-aware expected number of planets in the window, ${\mathbb E}[N_{\rm win}]=19.97\pm2.77$ (total) and $18.68\pm2.60$ (ridge). These values correspond to deviations of $z=(N_{\rm obs}-{\mathbb E}[N_{\rm win}])/\sigma\simeq1.81$ (total) and $z\simeq2.04$ (ridge).

We also define a KDE-based bump-strength statistic, $S$, at $\rho_{\rm p}=1.7\,\mathrm{g\,cm^{-3}}$ as the logarithmic ratio between the KDE value at the target density and a local baseline estimated from symmetric sidebands. Using a non-parametric bootstrap, we obtain $S=0.364$ with a 68\% interval [0.116,\,0.582] for the total sample and $S=0.477$ with a 68\% interval [0.242,\,0.701] for the ridge subsample. By construction, $S>0$ indicates a positive bump relative to the local baseline.

\subsection{Unimodality diagnostics}

Finally, we apply Hartigan's dip test for unimodality. The resulting p-values are $p=0.936$ (total) and $p=0.290$ (ridge), indicating that unimodality cannot be rejected at high significance.

Overall, while formal unimodality tests do not require a fully separated bimodal structure, multiple independent diagnostics consistently reveal a statistically persistent local maximum near $\rho_{\rm p}\simeq1.7\,\mathrm{g\,cm^{-3}}$, particularly within the ridge population.

\end{appendix}

\end{document}